\pdfoutput=1
\documentclass[3p,twocolumn]{elsarticle}

\usepackage[cmex10]{amsmath}
\usepackage{amssymb}
\usepackage{extarrows}
\usepackage{graphicx}

\usepackage[nodots,nocompress]{numcompress}

\biboptions{comma,square}

\journal{Nano Communication Networks}

\newcommand{\pH}{\ensuremath{\mathrm{pH}}}

\begin{document}

\begin{frontmatter}



\title{Nano-scale reservoir computing}


\author[ICTC]{Oliver Obst\corref{cor1}}
\ead{oliver.obst@csiro.au}
\cortext[cor1]{Corresponding author. Tel.: +61 2 9372 4710 Fax: +61 2 9372 4310.}
\author[CMSE]{Adrian Trinchi}
\author[CMSE]{Simon G. Hardin}
\author[ICTC]{Matthew Chadwick}
\author[CMSE]{Ivan Cole}
\author[CMSE]{Tim H. Muster}
\author[CMSE]{Nigel Hoschke}
\author[ICTC]{Diet Ostry}
\author[CMSE]{Don Price}
\author[CMSE]{Khoa N. Pham}
\author[ICTC]{Tim Wark}

\address[ICTC]{Commonwealth Scientific and Industrial Research Organisation (CSIRO)\\
Computational Informatics, PO Box 76, Epping NSW 1710, Australia}
\address[CMSE]{Commonwealth Scientific and Industrial Research Organisation (CSIRO)\\
Materials Science \& Engineering, Private Bag 33, Clayton South Vic 3169, Australia}

\begin{abstract}
This work describes preliminary steps towards nano-scale reservoir computing using quantum dots. Our research has focused on the development of an accumulator-based sensing system that reacts to changes in the environment, as well as the development of a software simulation. The investigated systems generate nonlinear responses to inputs that make them suitable for a physical implementation of a neural network. This development will enable miniaturisation of the neurons to the molecular level, leading to a range of applications including monitoring of changes in materials or structures. The system is based around the optical properties of quantum dots. The paper will report on experimental work on systems using Cadmium Selenide (CdSe) quantum dots and on the various methods to render the systems sensitive to pH, redox potential or specific ion concentration. Once the quantum dot-based systems are rendered sensitive to these triggers they can provide a distributed array that can monitor and transmit information on changes within the material.
\end{abstract}


\end{frontmatter}


\section{Introduction}
In our work we investigate a combination of recent advances in materials science and information processing (see~\cite{BTB+11} for a discussion on the role of inter-disciplinary research in nanoscale communication). The goal of this research is to enable new classes of devices through the development of smart molecular systems incorporating nanoparticles serving in accumulator-based sensor applications. The sensing system that is being developed aims to mimic that of a neural network, in terms of its sensing and signal propagation. In this system, active particles with simple capabilities per device serve a triple function. First, as input units in a physical realisation of an artificial neural network, they can collect and transfer information from external stimuli. Second, in internal layers, they accumulate information from neighbouring regions. Third, and perhaps most critically, they are able to respond when the input exceeds a certain threshold.

From the information processing point of view, Reservoir Computing (RC) appears to be a suitable approach to achieve this goal (for a recent overview of this field, see~\cite{LJ09}). RC computing approaches have been employed as mathematical models for generic neural microcircuits, as well as to investigate and to explain computations in neocortical columns (see, e.g.,~\cite{MNM02}). A key element of RC approaches is the randomly constructed, fixed hidden layer -- typically, only connections to output units are trained.

The vision for the work described in this paper is to create a network of particle-sized units that collectively process information coming from the surface of a material. Such a technology would have a variety of applications, from monitoring the conditions of a structure (e.g., corrosion or stress), to responding to changes in the environment. Potentially, with many simple and low-power units connected to a large network, this technology could also perform generic computations similar to a processor, but instead of single operations executed at a high frequency make use of massive parallelism at lower frequencies. 

\begin{figure}[t]
\centering
\includegraphics[width=0.95\columnwidth]{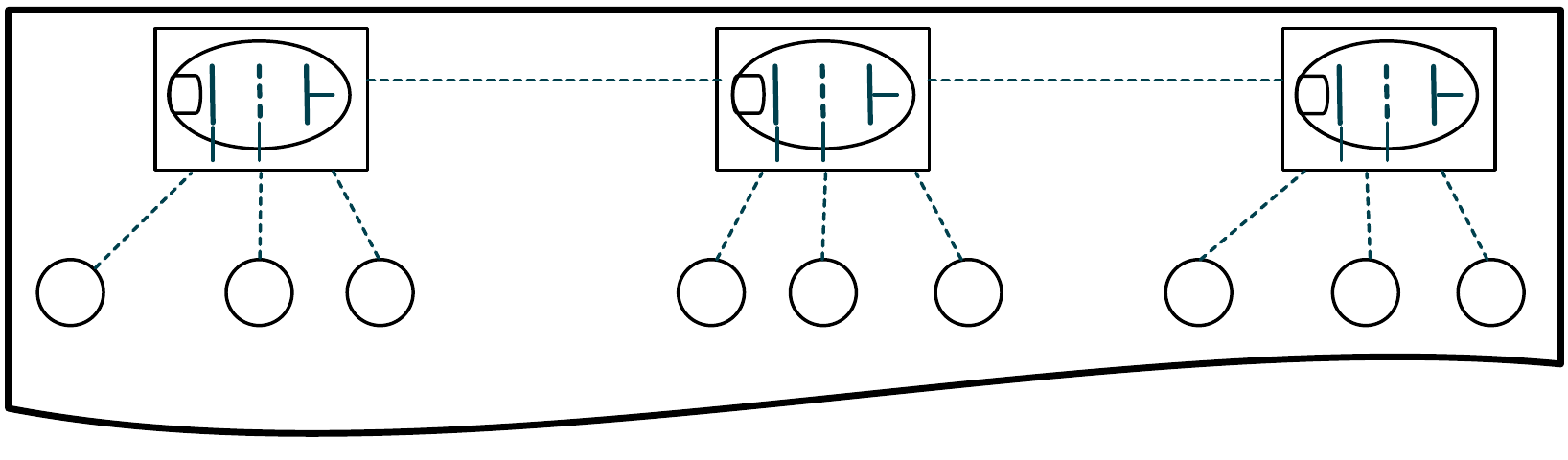}%
\caption{For a first working prototype, a network of sensors will be created using QDs as sensors, with help of nano-scale triodes.}
\label{fig:overview}
\end{figure}

We focus on quantum dots (QDs) as the devices to be used for sensing changing properties of a material. It may also be possible to get QDs to communicate signals in order to form a network. We describe the steps we have taken so far in creating an information-processing network of QDs in the subsequent sections. These steps include the materials science aspects of manufacturing the actual QDs, and also a simulation of processes in a material, as a testbed for further developments. Using this simulation, we can evaluate effects of changes in the material on the QDs attached to it. Using another software implementation, we simulate a nano-scale RC network to investigate the potential of recurrent neural networks with the topological constraints of an implementation in hardware. The physical implementation of this connectivity in hardware can be realised in various ways, and is subject of our ongoing research. A first, pragmatic approach is to aggregate groups of QDs, to treat these groups as a single sensors, and to pick their signals up using nano-scale triodes for further communication (see the schematic in Fig.~\ref{fig:overview}). We are currently working on such an approach, but are also planning to investigate optical or chemical signalling at a later stage. 

Section~\ref{sec:experimental} contains some details of the chemical processes involved in our quantum dot system. In essence, the idea is to use quantum dots to sense, to accumulate and transmit information from the surrounding environment. The sensory stage will be located physically close to external stimuli which may be in the solid, liquid or gaseous state. As a result of this interaction, a signal that may be chemical, electrochemical, thermal or optical in nature is propagated through the medium until it reaches the accumulating stage. These accumulating particles are affected by the signal in a characteristic way such that the accumulator is influenced by each signal, but does not undergo a change in the physical quantity being measured until a critical number of signals have reached it. Once this critical number of signals has been reached, the accumulator changes one of its properties and in doing so sends an intense signal to the responding particle; the latter either changes the properties of the structure or sends an amplified signal from the structure. Advantages over software based RC implementations would be speed, parallelism, as well as energy efficiency.

In Sect.~\ref{sec:models} we describe some aspects of our software simulation and some of the processes involved: the system that we use to experiment with possible configurations and network architectures, an abstract simulation of some of the chemical processes above in software.

Alternative physical implementations of RC have been proposed before. The work of Vandoorne et al.~\cite{VDV+11}, for example, proposes the use of coupled semiconductor optical amplifiers. These devices can be placed on a small chip and implement a number of units. A different approach, using electronic circuits, nano-wires and self-assembly, is presented in the paper of Stieg et al.~\cite{SAS+11}. Our approach uses essentially nanoparticle-sized units, so that it may allow an embedding of the entire reservoir network into, for example, the coating of a material.

\section{Experimental work using quantum dots}
\label{sec:experimental}

QDs are spherical submicron particles (typically 1-10nm), normally of a semi-conducting material, and often have a surrounding shell of a second semi-conducting phase. The optical properties of the QDs may be tailored by doping them with other elements or by particle size control. 
The most notable characteristic of QDs is their ability to absorb energy over a wide range whilst fluorescing with a relatively narrow bandwidth at a longer wavelength. One of the most widely investigated QD systems is Cadmium Selenide (CdSe) with a Zinc sulphide (ZnS) shell. These QDs absorb strongly through the mid- and near-UV and into the visible, yet emit a strong fluorescence signal in a relatively narrow band in the red region of the visible spectrum, with a high quantum efficiency. 

The focus of our experimental work thus far has been on the development of the accumulator-based sensing system, which utilises CdSe/ZnS QDs coupled with a signal conversion molecule that reacts to changes in the environment. The signal conversion molecules selected have been either pH-sensitive dye materials (Sect.~\ref{sec:ph}) or redox potential-sensitive dye materials (Sect.~\ref{sec:redox}). The accumulator consists of the quantum dots surrounded by a medium containing either the pH or redox sensitive dye molecules. These surrounding dyes either block the incoming light that stimulates the quantum dot fluorescence, or serve to absorb and hence shut down the QDs outgoing fluorescence. The 
QDs used in this work has an emission peak centred at 646nm (FWHM $\approx 25$nm).

\subsection{pH System}
\label{sec:ph}

The pH sensitive dye molecule selected for this demonstration was p-nitrophenol (4-hydroxy\-nitro\-benzene) as it possesses suitable optical properties for blocking the incoming optical stimulus to the quantum dots. P-nitrophenol is a weak acid with a $\mathrm pK_a$ of 7.08. In its neutral form, i.e., when the dye molecule is not an ion, at $\pH < 4$ the molecule is colourless, with transitions occurring in the UV near 300nm. At $\pH > 4$, the molecule progressively deprotonates at the phenolic group, and transitions near 400nm emerge, resulting in a yellow colour. This anionic form of the molecule has several resonance-stabilised forms, causing lower-energy transitions appear, compared to those of the neutral form. Above $\pH$ 7, the process is essentially complete. Thus progressive accumulation (or loss) of protons can be used as the mechanism in the accumulation stage of the system, where 400nm is used as the excitation wavelength. Figure~\ref{fig:absorption} below illustrates the change in the absorption spectrum of p-nitrophenol at different $\pH$ values, highlighting the optical transitions around 400nm.

\begin{figure}[ht]
\centering
\includegraphics[width=0.95\columnwidth]{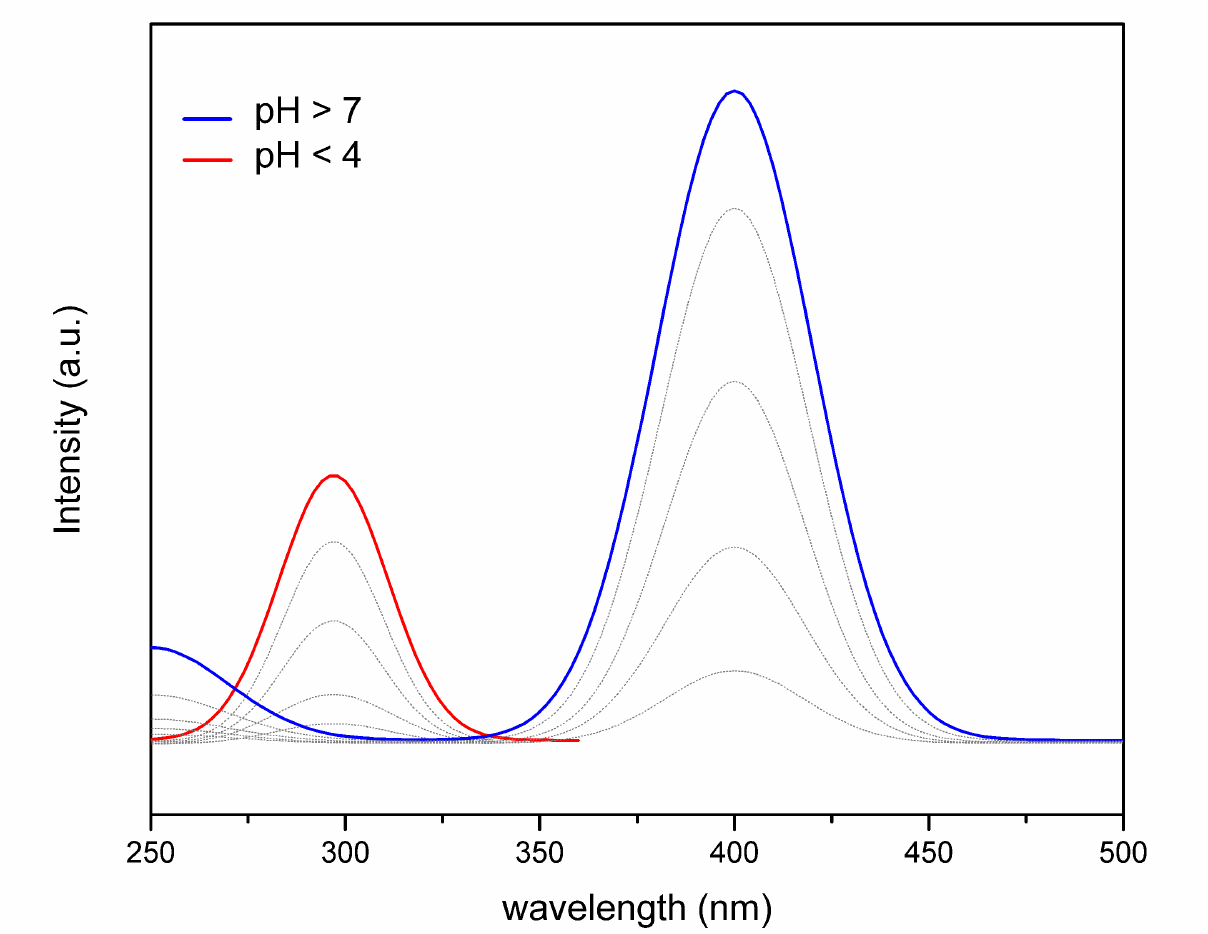}
\caption{Absorption spectra of p-nitrophenol at different $\pH$ levels.}
\label{fig:absorption}
\end{figure}

\begin{figure}[bt]
\centering
\includegraphics[width=0.85\columnwidth]{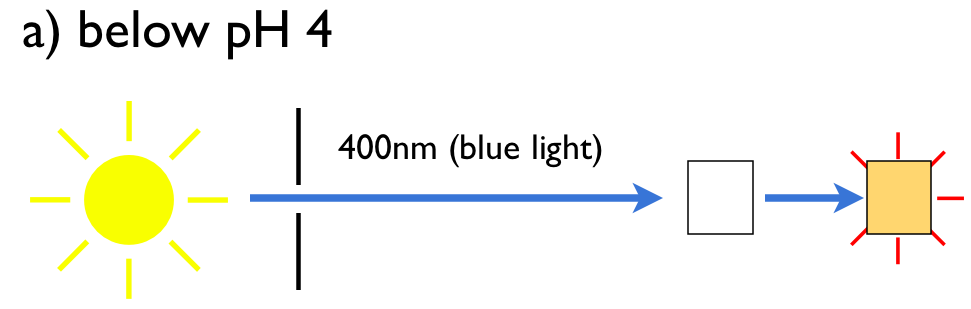}%
\hfill%
\includegraphics[width=0.85\columnwidth]{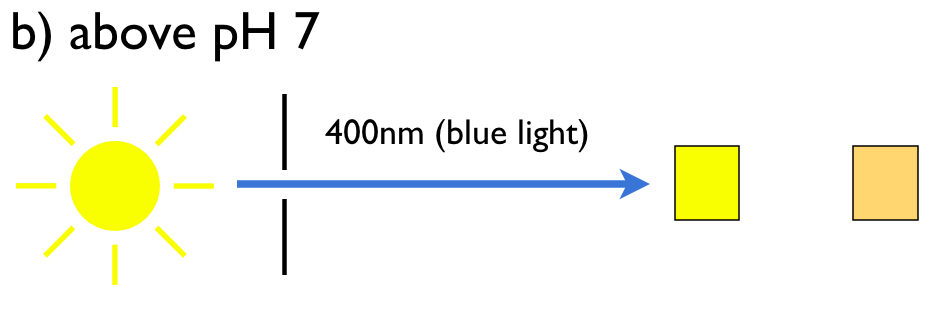}
\caption{Fluorescence test cell configuration.}
\label{fig:fluorescence}
\end{figure}

\begin{figure*}[ht]
\begin{minipage}[t]{\columnwidth}
\centering
\includegraphics[width=\textwidth]{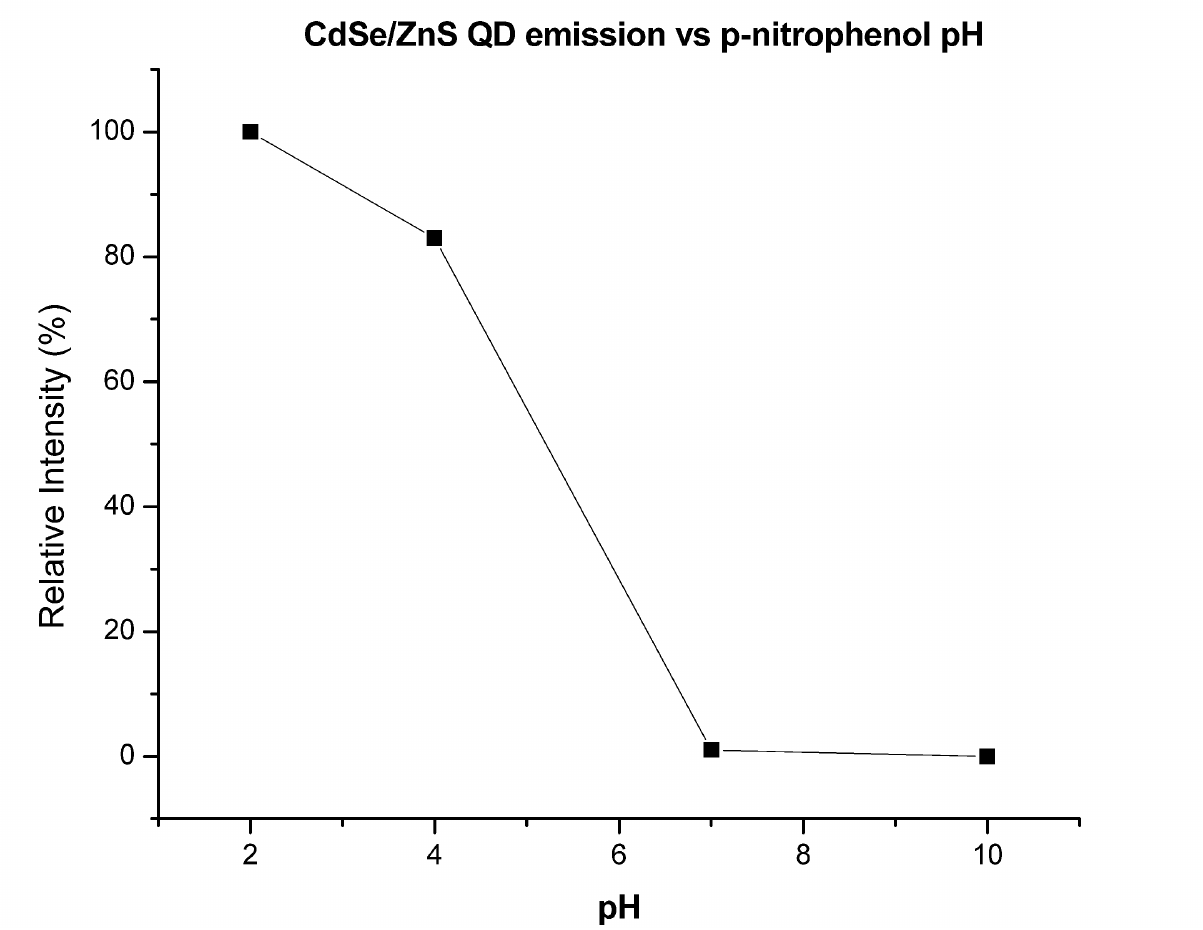}
\caption{QD emission vs.~p-nitrophoenol at different pH concentrations}
\label{fig:emission-vs-p}
\end{minipage}
\hfill
\begin{minipage}[t]{\columnwidth}
\centering
\includegraphics[width=\textwidth]{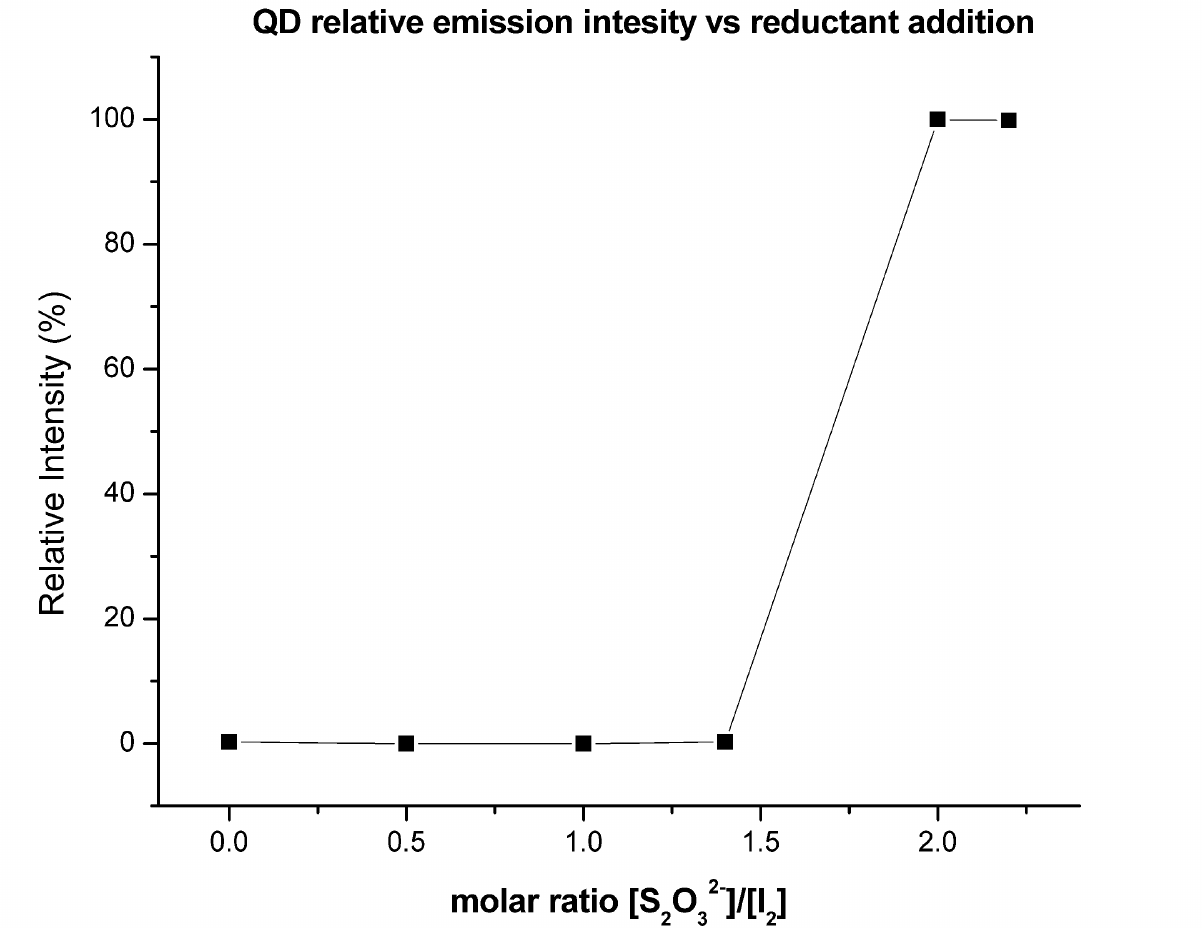}
\caption{QD emission vs.~added concentration of sodium thiosulphate}
\label{fig:emission}
\end{minipage}
\end{figure*}

A custom test cell configuration, illustrated by the schematic in Fig.~\ref{fig:fluorescence}, was utilised to prove the concept. In the first cell, a solution containing the pH sensitive dye p-nitrophenol was placed into the path of the incoming 400nm excitation light. The other test cell containing the quantum dot solution was then placed directly behind the first cell. Tests were carried out in Cary Eclipse fluorescence spectrophotometer. The pH of the p-nitrophenol solution was then systematically adjusted upwards by the use of buffers, and the fluorescence emission of the QDs monitored. As can be seen in Fig.~\ref{fig:emission-vs-p}, it is quite clear that p-nitrophenol is able to completely absorb the incoming 400nm radiation at $\pH$ 7 or above, preventing the QDs from fluorescing. At $\pH < 4$, the neutral molecule is transparent at 400nm, allowing this radiation to be absorbed by the QDs, with subsequent fluorescence.

\subsection{Redox system}
\label{sec:redox}

In a second variation of the accumulator concept, a redox system was used to block the outgoing fluorescence of the QD. In a redox system, electrons may be gained or lost due to reduction or oxidation, respectively. The system used was NaI/NaI$_3$/Na starch glycolate. Neutral I$_2$ has a low solubility in water, but dissolves in excess iodide (e.g. NaI) to give triiodide, I$_{3^-}$, a soluble species. Both neutral iodine and the anion give an intense blue colour with starch and its derivatives due to the formation of a molecular addition compound. The absorption has a maximum near 600nm, with strong absorption at 646nm. On the other hand, no coloured addition compound is formed with I$^-$ and starch. Iodine and iodine are easily interconverted with standard oxidising or reducing agents, or by application of a potential. When in the presence of reducing agent, the solution becomes colourless due to the I$_{3^-}$ ions being reduced to three iodide ions, causing the disappearance of the deep blue starch-iodine complex.

\begin{figure}[tb]
\centering
\includegraphics[width=0.9\columnwidth]{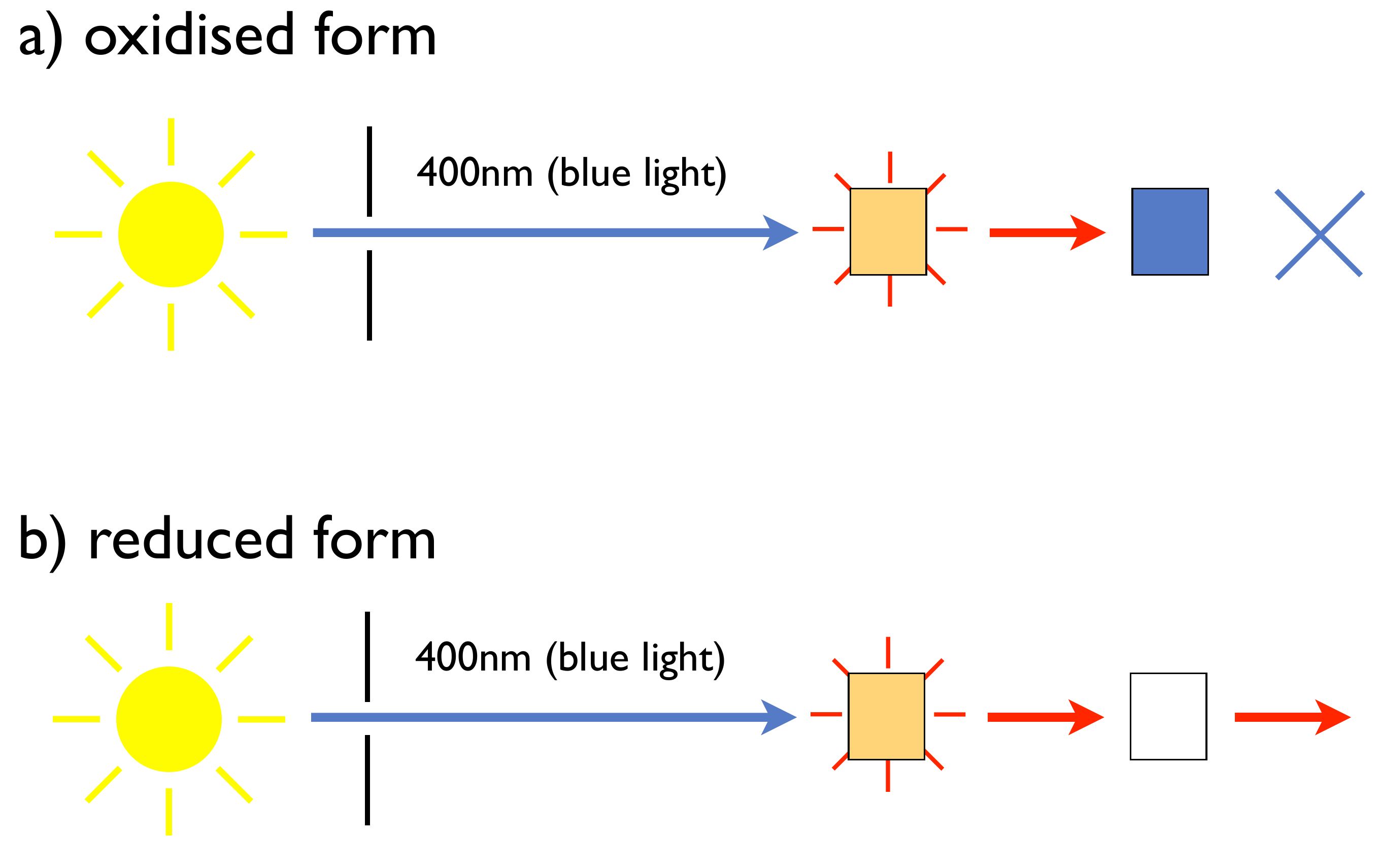}%
\caption{Redox test cell configuration.}
\label{fig:redox}
\end{figure}

Half reactions are:
\begin{align}
2\mathrm{I}^- & \longrightarrow \mathrm{I}_2 + 2e^-\\
\mathrm{I}_2 + \mathrm{I}^- & \longrightarrow \mathrm{I}_3^-,
\end{align}
yielding:
\begin{align}
\mathrm{I}_3^- & \xLongleftrightarrow{\pm 2 e^-} 3\mathrm{I}^-
\end{align}
Chemically, reduction was accomplished by addition of thiosulphate $\mathrm{(S_2O_3)^{2-}}$ ions. Experimentally this was achieved by placing one cell containing the $\mathrm{NaI/NaI_3/Na}$ starch glycolate behind the QD-containing cell, at right angles to the incident beam, i.e., in the direction of fluorescence (see Fig.~\ref{fig:redox}). Subsequently, progressive amounts of a reducing agent, sodium thiosulphate $\mathrm{(Na_2S_2O_3)}$, were added, being a standard reagent for iodine reduction. The reduction was carried out according to:
\begin{align}
	\mathrm{2(S_2O_3)^{2-} + I_2} & \rightarrow \mathrm{(S_4O_6)^{2-} + 2I^-}
\end{align}
During the experiments, an excitation wavelength of 400nm was again used. It was found that the QD fluorescence was completely absorbed until virtually all the $\mathrm{I_{3^-}}$ had been reduced to $I^-$, demonstrating an accumulator effect in the presence of electron donation (i.e., reduction, see Fig.~\ref{fig:emission}). In a preliminary investigation, we have also demonstrated electrochemical generation of colour in the same system.

\section{Software-based simulation}
\label{sec:models}

To experiment with possible configurations and network architectures, we implemented an abstract simulation of some of the processes above in software. Our Cellular Automata (CA) based models consist of coupled 2d layers, each simulating a different aspect of the system.

\subsection{The chemical model}

The lower layers of our simulation have the goal to serve as sample input to processes in higher levels of our model.  We use a simple layered 2d cellular automaton, where each cell in a layer is a 4-tuple $(x_1, x_2, x_3, x_4)$, $x_i \in [0..1]$.  Our chemical model is comprising of two layers. The first one serves is a chaotic process, currently the coupled logistic map:
($x_i \rightarrow a R x_i (1-x_i) + (1-a) R x_{i+1} (1-x_{i+1})$ with a high $R$ and low $a$ (we used $R = 4$, and both $a = 0.0$ and $a = 0.001$). This layer serves as a source of chaos for the next layer. The second layer is a simple electrochemical model of 4 parameters:
\begin{itemize}
\item F is a feature with a probability of inducing an anodic potential difference F+ or a cathodic potential difference F-.

\item V is a potential relative to the OCP (open circuit potential). V+ is
anodic to OCP, V- is cathodic to OCP.

\item pH is the pH with pHa as acidic and pHb as basic. 

\item Mz+ is the metal ion concentration. 
\end{itemize}

The probabilistic system progression is from $F+ \rightarrow V+ \rightarrow Mz+ \rightarrow pHa \rightarrow F+$ and from $F- \rightarrow V- \rightarrow pHb \rightarrow F-$.

The state of a point $i$ in the 2d space is given by a vector $\mathbf{a}^i$, 
with $a^i_1 = \mathrm{F}$, $a^i_2 = \mathrm{V}$, $a^i_3 = \mathrm{pH}$, and $a^i_4 = \mathrm{Mz+}$, respectively. This layer updates each cell by summing the
concentrations of these values over its neighbours $N$, if values from the chaotic process in the layer below, interpreted as transition probabilities $\mathbf{p}$, exceed a given threshold $\mathbf{\theta}$. 

\begin{align}
\mathbf{a}^\mu & = \frac{1}{1+|N|} (\mathbf{a}^i + \sum_{\mathbf{a}^k \in N} \mathbf{a}^k)\\
{a}^i_2 & = {a}^\mu_1, \text{\qquad if $p_1 > \theta_1$} \\
{a}^i_3 & = {a}^\mu_2, \text{\qquad if $p_2 > \theta_2$} \\
{a}^i_4 & = {a}^\mu_3, \text{\qquad if $p_3 > \theta_3$} \\
{a}^i_1 & = {a}^\mu_4, \text{\qquad if $p_4 > \theta_4$ and $a^\mu_4 > 0$}
\end{align}

\subsection{The quantum dot model}

A simple model of quantum dots is used as a layer on top of the chemical model. In this layer, each cell (again, a 4-tuple) is excited in each of its components by each of the concentrations of parameters in the electrochemical layer below (i.e., in our model taking values from one of the vectors $\mathbf{a}^i$ above), plus some excitation from its immediate neighbours (the $\mathbf{a}^k \in N$ above). Therefore each cell is assumed to hold 4 quantum dots, represented as red, green, blue and alpha, which in turn represent fluorescence due to F, V, M and pH in layer 2.

\begin{figure}[tb]
\centering
\includegraphics[width=0.47\columnwidth]{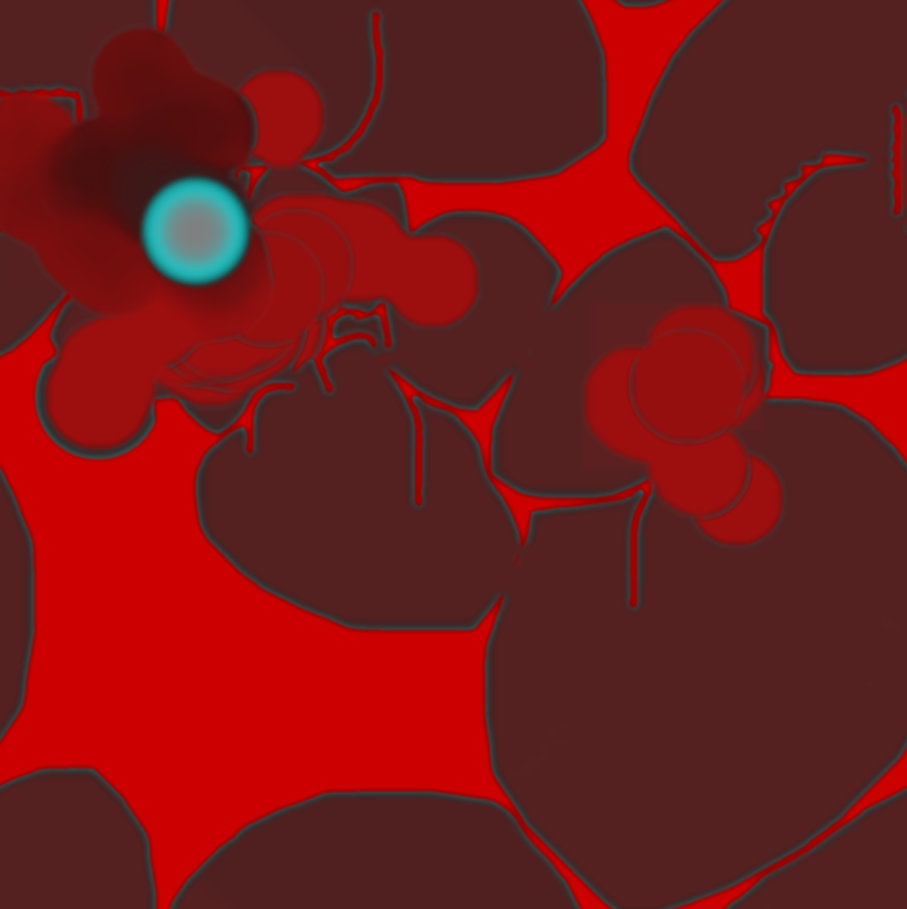}\hfill 
\includegraphics[width=0.47\columnwidth]{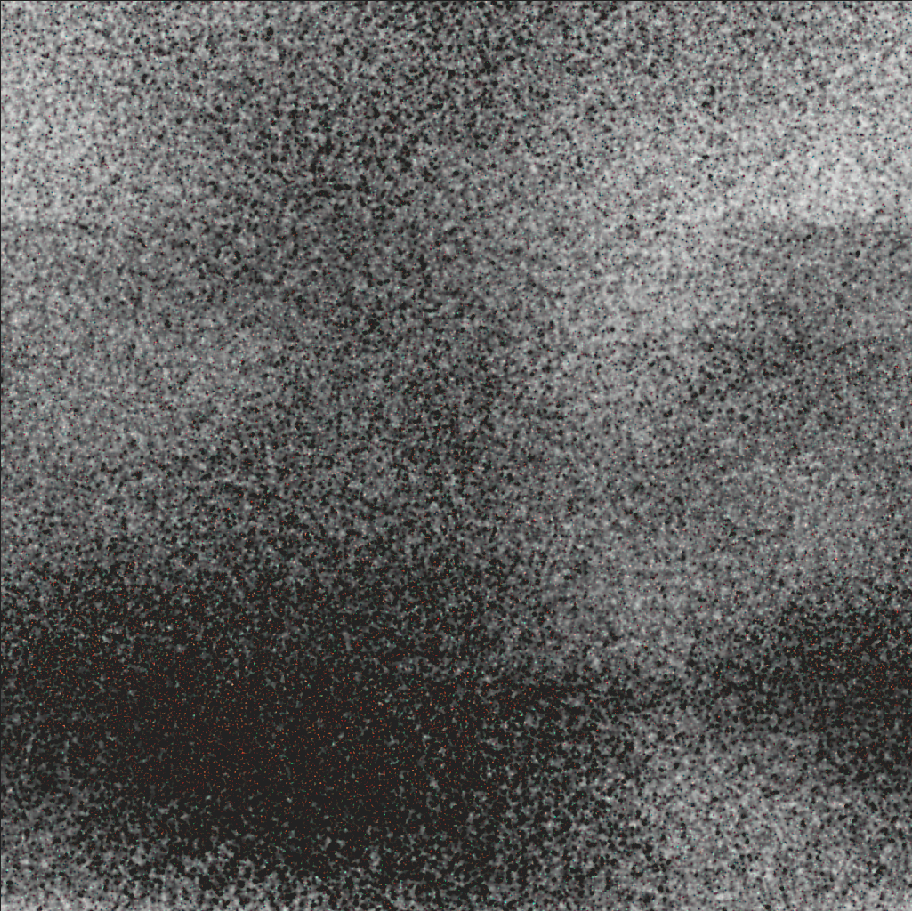} 
\caption{Different versions of Cellular Automata (CAs), used for simulation of an abstract chemical process. The left screen shot is from a Gray-Scott model, the right one using the model described in the text as an abstraction of a corrosion process.}
\label{fig:ca}
\end{figure}

Using cellular automata, we can simulate abstract chemical processes (see Fig.~\ref{fig:ca} for example screenshots), that may be monitored using a coating with an embedded network of QDs. One approach to classify inputs using randomly placed elements is described in the following subsection.

\subsection{Reservoir computing}

Recurrent neural networks (RNN) are one approach to model dynamical systems, and produce outputs based on some input and their current internal state. They can be used for prediction or classification, provided the connections weights between neurons are appropriately set by a training procedure. Traditional RNN training methods can suffer from problems like slow convergence and vanishing gradients~\cite{Hoc91,BSF94}. 

To address these challenges, a mathematical model for generic neural microcircuits, the \emph{liquid state machine} (LSM) was proposed~\cite{MNM02}. The framework for this model is based on real-time computation without stable attractors. The neural microcircuits are considered as dynamical systems, and the time-varying input is seen as a perturbation to the state of the high-dimensional excitable medium implemented by the microcircuit. The neurons act as a series of non-linear filters, which transform the input stream into high-dimensional space. These transient internal states are then transformed into stable target outputs by readout neurons, which are easy to train. This approach to neural modeling has become known as \emph{reservoir computing} (see also, e.g.,~\cite{LJ09}), and the LSM is one particular kind of model following this paradigm.

\begin{figure}[btp]
\centering
\includegraphics[width=0.74\columnwidth]{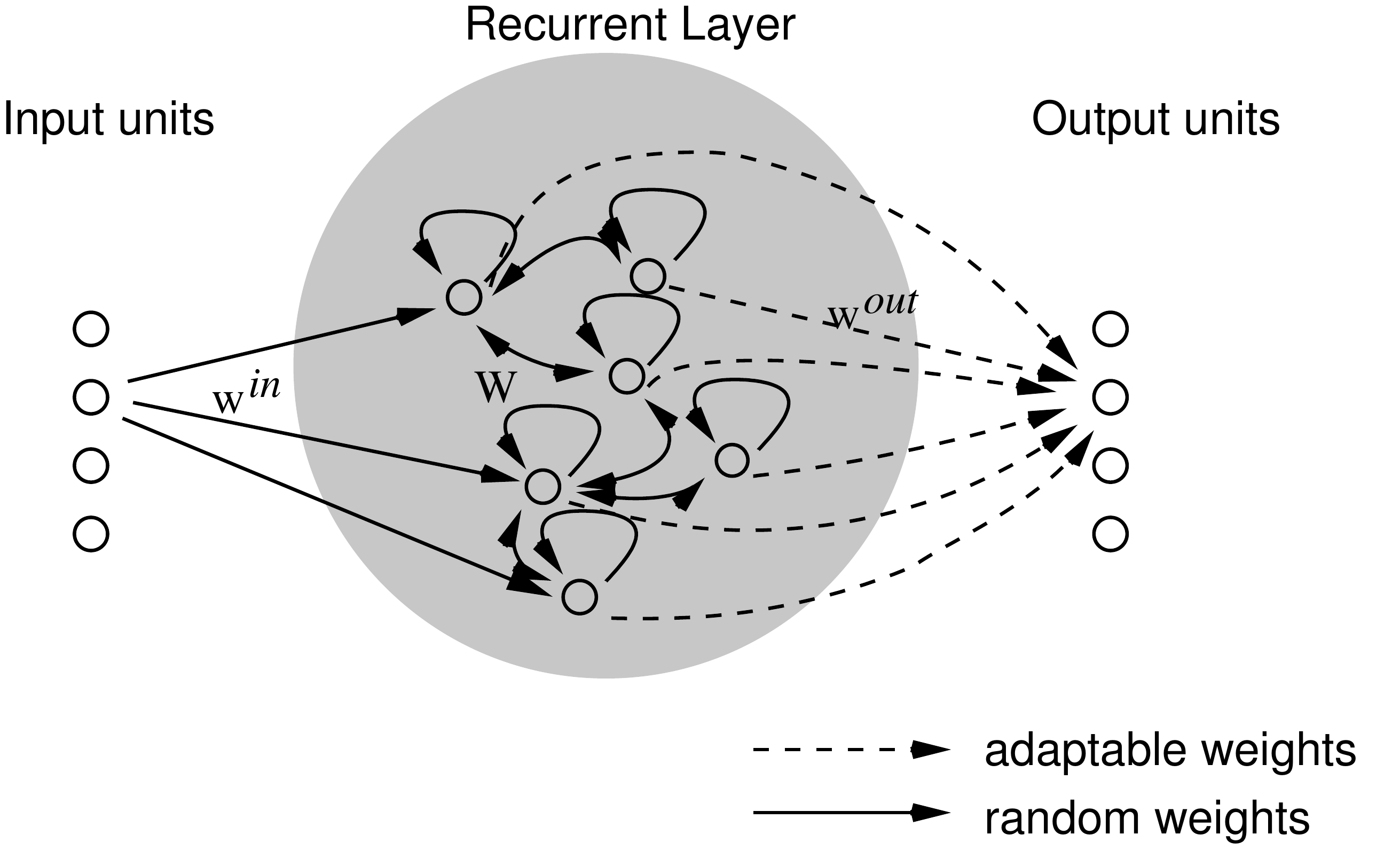}
\caption{Architecture of an echo state network. In ESNs, usually only the connections represented by
the dashed lines are trained, all other connections are setup randomly and remain fixed. The recurrent layer is also called a \textit{reservoir}, analogously to a liquid, which has fading memory properties.}
\label{fig:esnscheme}
\end{figure}

\emph{Echo state networks} (ESN)~\cite{JH04} are a reservoir computing model similar 
to LSM. They implement the same concept of keeping a fixed high-dimensional reservoir of neurons, with random connection weights between reservoir neurons small enough to guarantee stability. Learning procedures train only the output weights of the network, but while LSM use spiking neuron models, ESN are usually implemented with sigmoidal nodes, which are updated in discrete time steps. An illustration of an ESN architecture is shown in Fig.~\ref{fig:esnscheme}. A reservoir-computing inspired approach is interesting for an implementation with QDs, because hidden connections do not need to be trained, allowing for random placements of QDs.

\subsection{Simulation of reservoir computing approaches in hardware}

In a software implementation of neural networks, no costs are associated with connecting any two arbitrary neurons. Reservoir computing approaches typically make use of this by fully connecting layers.
In a hardware implementation, each neuron has a spatial location, 
resulting in a cost to connect to another neuron (e.g., wiring length), and some connections may not be feasible at all. 
ESNs with a one-dimensional topography have been investigated in~\cite{MBW10}, and a topography where several (randomly connected) reservoirs are regularly connected in a grid-like fashion using a wireless sensor network is used in our previous work~\cite{Obst09a}. Despite these constraints on connectivity between hidden units, 
e.g., prediction of future states is possible with a reservoir computing method -- in the case of sensor networks with one of the applications to detect anomalies~\cite{JWOV11}.
Use of piecewise linear approximations of the hyperbolic tangent transfer function in individual units, similar to our nonlinearities in 
 Figs.~\ref{fig:emission-vs-p} and \ref{fig:emission}, have been investigated in~\cite{CTB09}.
Some other restricted topologies have also been a subject of investigation in~\cite{VDV+11}.

\begin{figure}[btp]
\centering
\includegraphics[width=0.47\columnwidth]{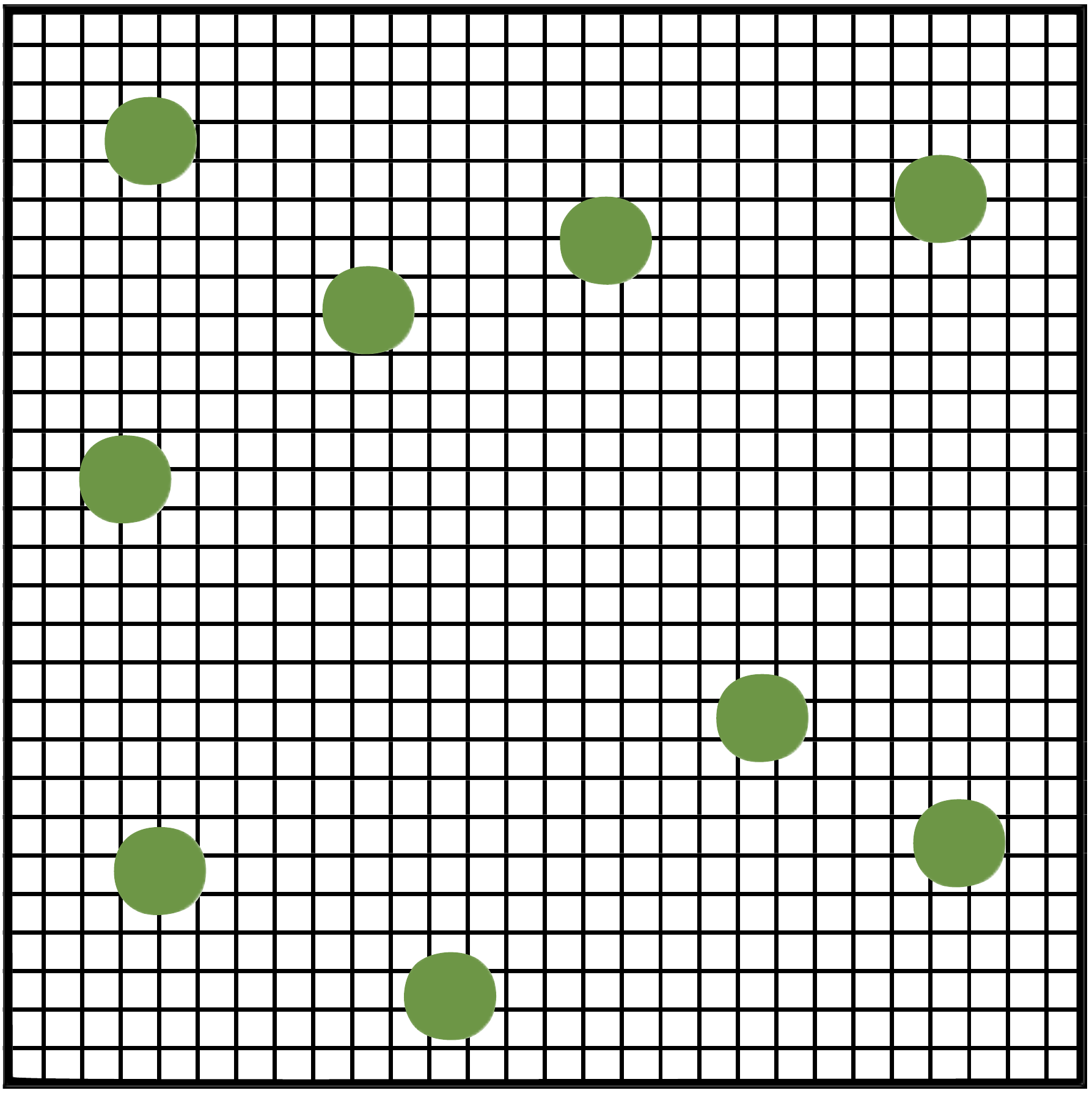} \hfill
\includegraphics[width=0.47\columnwidth]{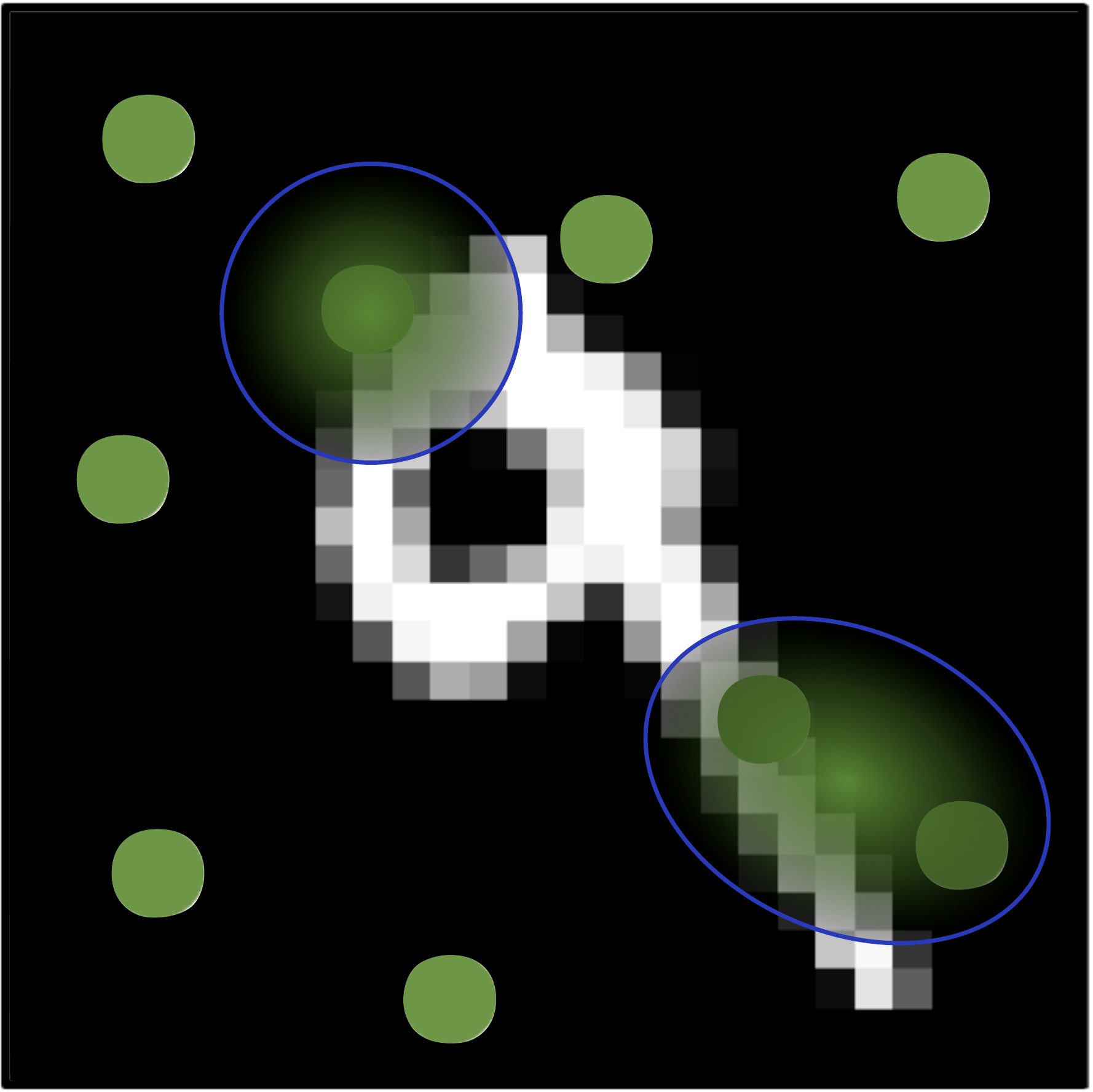}
\caption{(Left) Randomly placed neurons over a grid of $28\times28$ inputs.\qquad (Right) Input pattern presented on the surface; neurons are activated both from the input pattern and from nearby active neurons.}
\label{fig:2dNNschema}
\end{figure}

In the following, we use a simple setup of a reservoir with a two-dimensional topography to demonstrate feasibility of a restricted connectivity. For this demonstration, we restrict the input-to-hidden and hidden-to-hidden layer connectivity, so that strength of  connections between two neurons falls off exponentially with their distance from each other, i.e., we get
\begin{align}
W^\text{in}_{i,k} & = \exp({\frac{-\mathrm{dist}(i,k)}{2 \mu}}), \label{eq:win} \\
W_{i,j} & = \exp({\frac{-\mathrm{dist}(i,j)}{2 \mu}}) + \epsilon_{i,j}, \label{eq:w}
\end{align}
for connections between input unit $k$ and hidden unit $i$, or connections between hidden units $i$ and $j$, respectively. A small constant $\mu$ (e.g., $\mu = 0.01$) is used to adjust the falloff of signal propagating from a neuron. $\epsilon_{i,j}$ is normally distributed noise with zero mean, and small standard deviation (e.g., $\sigma = 0.007$). With this small amount of noise, the resulting connection matrices will be nearly symmetric since connection strength is a function of distance. For our demonstration, we use a regular $28\times28$ grid representing input, e.g., from some surface area of a material. Objects in our experimental data set are represented by $28\times28$ pixels, but the exact size is not a limitation of our approach, and can be adjusted.
Over this area, we randomly distribute 100 neurons that are connected to the input and to each other according to (\ref{eq:win}) and (\ref{eq:w}), respectively. These neurons become active when inputs in their proximity are active, and also when other neurons in their neighbourhood are active (see also the schemata in Fig.~\ref{fig:2dNNschema}). Smaller (and also larger) number of neurons can be chosen. Activation of the neurons $\mathbf X$ are updated based on the states of their neighbours and inputs $\mathbf I$:
\begin{align}
{\mathbf X}_{t+1} = \tanh({\mathbf W \mathbf X}_{t} + {\mathbf W}^\text{in} \mathbf{I}_{t+1})
\end{align}
Connected to the hidden layer, we have a set of 10 outputs. 
Connection strengths to these outputs is determined by a training procedure. 
For this demonstration, output units may be fully connected. 
During the training, we present 60000 handwritten digits from the MNIST dataset~\cite{LBBH98} as an input to the system. 
Each input pattern is presented for a brief period until the state of the reservoir remains stable. Then, its state is collected in a matrix $\mathbf M$ ($100~\text{neurons} \times 60000$ steps), in addition to a matrix $\mathbf V$ containing the desired state of the 10 outputs for each of the pattern ($10 \times 60000$). 
Matrices $\mathbf{M}$ and $\mathbf{V}$ can now be used to calculate the desired weights from hidden layer to the outputs, by using simple linear regression, using the Moore-Penrose pseudoinverse of $\mathbf M$, i.e.,
\begin{align}
\mathbf{W}^\text{out} & = \mathbf{V} \mathbf{M}^{+}.
\end{align}

Alternatively, a logistic regression can also be used, and will automatically restrict the outputs to values.  
When we present a new pattern to the system, we can classify the input into one of 10 classes by selecting the most active output. For a test with 10000 new patterns, this approach classified approx.~$85\%$ using linear regression or $91\%$ (logistic regression) of the presented handwritten digits correctly. The purpose of this experiment is not to achieve a competitive recognition rate -- even simple, but spatially unconstrained approaches achieve much higher rates on the very same data set -- but to show that even with these constraints complex computation may be possible. 

For a system implemented with quantum dots, the number of neurons in the hidden layer could be much larger. 

\section{Conclusion}

In this work, we have demonstrated the feasibility of the accumulator-based sensing system. This constitutes first steps towards a nano-scale reservoir computing approach. On the chemistry side, our next steps will be to miniaturise the accumulator to the molecular level. An investigation is currently underway to determine suitable approaches for fabricating them. The first attempt currently undertaken involves coupling the QD particles with a pH sensitive dye. Approaches include incorporation of the materials into permeable polymeric films or in mesoporous inorganic particles. Research into suitable porous membranes, chromogenic polymers and growth strategies are currently underway. On the information processing side, we are investigating different connectivities as well as possible training methods~\cite{Obst09d}. One goal is to also constrain the number of connections to output units, for example by using hierarchies of hidden neurons. 

In principle, a system built using our approach may allow for universal computation. A more direct application, however, would be in monitoring of contaminants in fluids at high sensitivity. In particular, important areas include monitoring water quality, detection of hydrocarbons in petroleum exploration, detecting the presence of low levels of contamination in process gases or in bio-security areas. A second practical application is in 
detecting contaminants or changes in solid-state structures. This is of high importance for structural health monitoring of both engineering and bio-mechanical structures.
The third application of potential high impact is in the development of structures that respond to changes in their external environment. Such use has the potential to create new possibilities in ``ageless structures'', such as membranes for both water and hydrogen permeability, amongst others.

For the last two application areas, the structure is itself critical and could be a coating on a metal structure (aircraft etc.) or a lining on a bio-implant.












\end{document}